\documentclass{article} 
\usepackage{MLDD_workshop_2023,times}

\usepackage{amsmath,amsfonts,bm}

\def\eqref#1{equation~\ref{#1}}

\def\1{\bm{1}}

\DeclareMathAlphabet{\mathsfit}{\encodingdefault}{\sfdefault}{m}{sl}
\SetMathAlphabet{\mathsfit}{bold}{\encodingdefault}{\sfdefault}{bx}{n}

\usepackage{hyperref}
\usepackage{url}

\usepackage[utf8]{inputenc} 
\usepackage[T1]{fontenc}    
\usepackage{hyperref}       
\usepackage{url}            
\usepackage{booktabs}       
\usepackage{amsfonts}       
\usepackage{nicefrac}       
\usepackage{microtype}      
\usepackage{xcolor}         
\usepackage{authblk}
\usepackage{multirow}       
\usepackage{graphicx}
\usepackage{listings}

\usepackage{bbm}

\title{FlexVDW: A machine learning approach to account for protein flexibility in ligand docking}

\author{Patricia Suriana, Joseph M. Paggi, Ron O. Dror \\
Department of Computer Science \\
Stanford University \\
\texttt{psuriana@stanford.edu, jpaggi@stanford.edu, rondror@cs.stanford.edu} \\
}

\iclrfinalcopy 
\begin{document}

\maketitle

\begin{abstract}
Most widely used ligand docking methods assume a rigid protein structure. This leads to problems when the structure of the target protein deforms upon ligand binding. In particular, the ligand’s true binding pose is often scored very unfavorably due to apparent clashes between ligand and protein atoms, which lead to extremely high values of the calculated van der Waals energy term. Traditionally, this problem has been addressed by explicitly searching for receptor conformations to account for the flexibility of the receptor in ligand binding. Here we present a deep learning model trained to take receptor flexibility into account implicitly when predicting van der Waals energy. We show that incorporating this machine-learned energy term into a state-of-the-art physics-based scoring function improves small molecule ligand pose prediction results in cases with substantial protein deformation, without degrading performance in cases with minimal protein deformation. This work demonstrates the feasibility of learning effects of protein flexibility on ligand binding without explicitly modeling changes in protein structure.
\end{abstract}

\section{Introduction}
A critical problem in rational drug discovery is prediction of the position, orientation, and conformation of a ligand (e.g., a drug candidate) when bound to a target protein---i.e., the ligand's "binding pose." Protein-ligand docking methods, which are used to predict ligand binding poses, are key tools in drug discovery and molecular modeling applications \citep{kitchen2004docking,ferreira2015molecular}. 

\begin{figure}[h]
\centering
\includegraphics[width=0.6\textwidth]{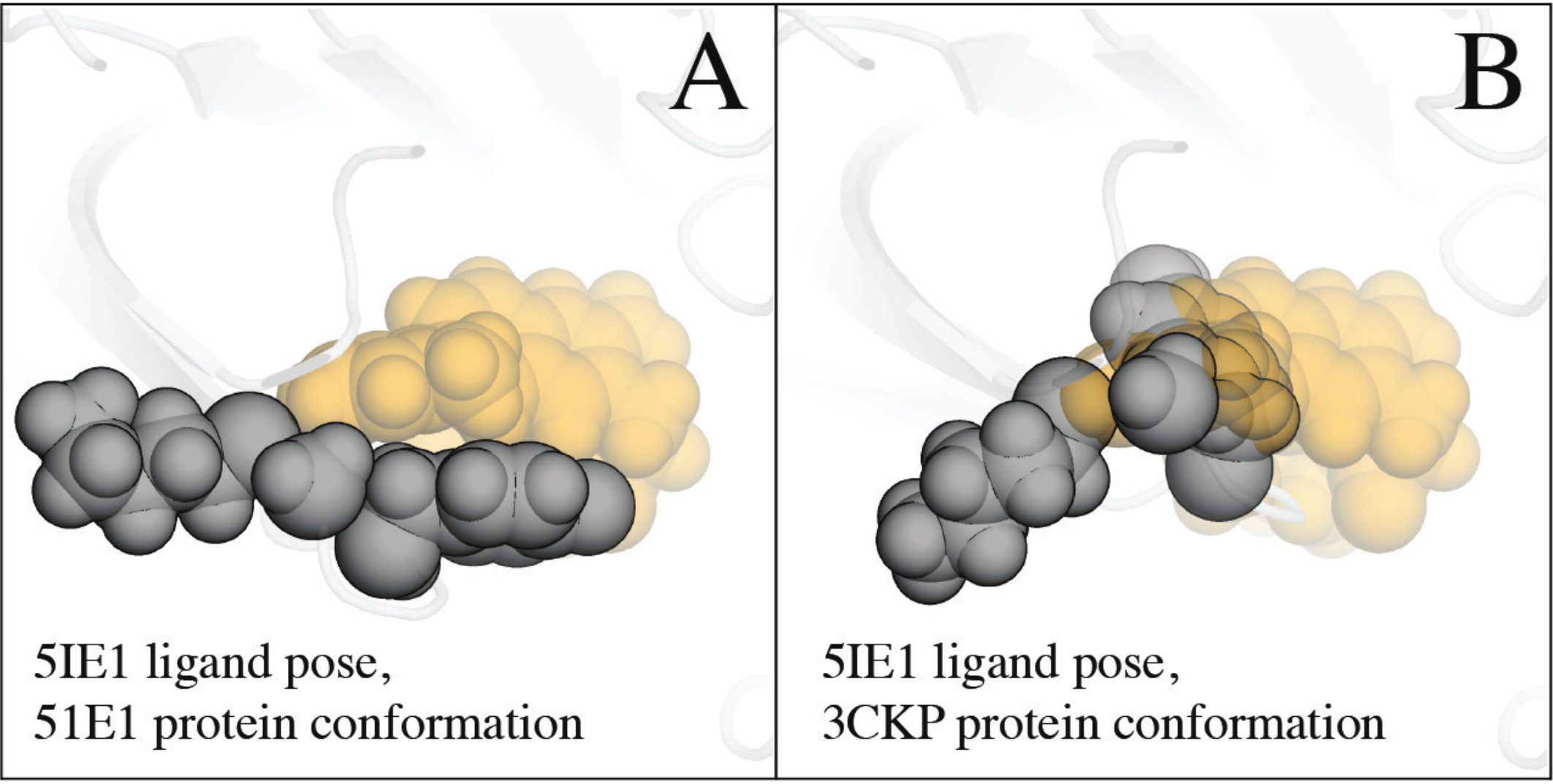}
\caption{The native binding pose of a ligand often clashes with the experimentally determined structure of its target protein when that structure has a different ligand bound. Panel A shows the structure of the protein $\beta$-secretase (BACE-1) ---- a major drug target --- bound to a ligand known as "compound 5" (PDB entry 5IE1 \citep{jordan2016fragment}). The ligand (orange spheres, with each sphere representing one atom) packs favorably against two amino acids in the protein binding pocket (gray spheres) without any clashes (i.e., ligand atoms do not overlap with protein atoms). Panel B shows the same ligand (compound 5) in exactly the same geometry, but superimposed on a structure of BACE-1 that was determined in the presence of a different ligand (PDB entry 3CKP, \citep{park2008synthesis}). Here, the same two amino acids (gray spheres) assume different positions and therefore clash (overlap) substantially with the ligand atoms (orange spheres).}
\label{fig:native_holo}
\end{figure}

The most widely used protein-ligand docking techniques assume a rigid protein (i.e., the positions of all protein atoms are fixed), which is often referred to as "rigid docking" \citep{verdonk2003improved,friesner2004glide,allen2015dock,forli2016computational}. Although this assumption of a rigid protein often works, rigid docking often fails to produce a near-native ligand pose (i.e., one that is close to the experimentally determined, or native, pose) when the shape of the protein's binding pocket must change for the ligand to bind. In such cases, atoms of the ligand in its native pose typically overlap ("clash") with atoms in the protein structure used for docking (Figure \ref{fig:native_holo}). Atoms that overlap experience extremely strong van der Waals repulsion. Rigid ligand docking methods thus predict that such poses will be extremely unfavorable energetically and generally rank them lower than any pose without such clashes---even when the clashes could have been easily resolved by minor changes in the structure of the protein's binding pocket. Such cases occur frequently in drug discovery, particularly when one is investigating novel ligands that differ substantially from ligands present in experimentally determined protein structures.

A variety of flexible protein docking techniques attempt to solve this problem by allowing the protein's binding pocket to deform during docking \citep{jones1997development,lemmon2012rosetta,miller2021reliable}. This approach is very computationally intensive, however, and has sometimes proven less accurate than rigid docking \citep{ravindranath2015autodockfr,bender2021practical}. Likewise, ensemble docking techniques in which each ligand is docked to multiple protein structures have met with mixed success, as selecting the protein structures and determining their relative favorability has proven difficult \citep{totrov2008flexible, novoa2010ensemble, amaro2018ensemble, evangelista2019ensemble,korb2012potential}.

In this work, we explore an alternative approach: rigid docking with a scoring function that has been adapted, through machine learning, to implicitly account for protein flexibility. In particular, we use an end-to-end machine learning approach to design a predictor of protein-ligand van der Waals (VDW) interaction energies. Given a single protein structure, our predictor is trained to recognize which types of deformations the protein's binding pocket can easily undergo, and to distinguish those from less favorable deformations. We name our machine-learned predictor of VDW interaction energies \textsc{FlexVDW}. 

We show that incorporating \textsc{FlexVDW} into an industry-standard docking package (Glide) improves ligand binding pose prediction results in cases where ligand binding requires significant protein deformation, without compromising performance in cases with minimal protein deformation. Our work demonstrates the feasibility of learning effects of protein flexibility on ligand binding without explicitly modeling changes in protein structure.


\section{Related Works}
In general, protein-ligand docking involves two challenges \citep{elokely2013docking, guedes2014receptor}: (1) designing a sampling algorithm to generate a large number of candidate ligand poses given a query ligand (i.e. ligand of interest) and a target protein structure, at least one of which should be close to the experimentally determined pose, and (2) designing a scoring function that ranks these candidate poses to select the best ones (i.e., those predicted to be closest to the native pose) as the final output. In this paper, we focus on the scoring challenge. Protein-ligand docking scoring function can be loosely categorized into two classes: (1) physics-based scoring functions, and (2) machine-learning scoring functions. 

Physics-based scoring functions \citep{friesner2004glide,verdonk2003improved,trott2010autodock,coleman2013ligand,allen2015dock} characterize the binding of protein-ligand complexes based on a set of weighted scoring terms, which correspond  to different physical effects such as VDW interactions, electrostatic interactions, hydrophobic interactions, and hydrogen bonds. The weights of these terms are typically determined by fitting experimental data using linear regression. It should be emphasized that the terms were carefully engineered to capture effects known to be important in determining ligand binding energy and represent decades of work.

Machine learning (ML) scoring functions \citep{khamis2015machine} allow for a more general functional form. Progress has been made in these areas, including end-to-end learning without hand-crafted features using deep learning methods \citep{shen2020machine,ragoza2017protein,morrone2020combining,mcnutt2021gnina}. Nevertheless, physics-based scoring functions such as Glide \citep{friesner2004glide} or DOCK \citep{coleman2013ligand,allen2015dock} have proven to be more generalizable to different drug target families, and especially to new drug targets not present in the training set, than ML-based functions and remain most widely used in drug discovery \citep{bender2021practical}. 
\section{Methods}

\subsection{Incorporating implicit protein flexibility into the scoring function for ligand docking}

Our goal in this work is to demonstrate the feasibility of creating a VDW interaction energy predictor that implicitly accounts for protein flexibility. We therefore develop a neural network, \textsc{FlexVDW}, that predicts VDW interaction energy. To demonstrate its effectiveness, we integrate \textsc{FlexVDW} into Glide \citep{friesner2004glide}, which is among the most widely used protein-ligand docking packages in the pharmaceutical industry. In particular, we replace the VDW term in Glide's physics-based scoring function with \textsc{FlexVDW}. Although in this work we chose Glide to incorporate our machine-learned VDW term, in principle our approach can be integrated with any existing physics-based scoring function, not limited to Glide, with refitting to the particular physics-based scoring function of interest. 

When training our neural network (but not when using it to predict ligand docking poses), we take advantage of the fact that, for certain proteins, multiple experimentally determined structures are available, with a different ligand bound to the same protein in each structure. Adopting terminology from structural biology, We refer to each of these ligand-bound structures as a "holo" structure. The set of holo structures for a given protein captures multiple shapes the protein's binding pocket can adopt and thus provides information about the binding pocket's flexibility. 

More concretely, the input to our ML model is a single protein structure to be used for docking a ligand, where the protein structure was determined in the presence of a different ligand, or with no ligand present at all. Our training labels, on the other hand, are generated by taking into account all available holo structures (see Figure \ref{supfig:pipeline}). Importantly, our model can be used to predicting ligand binding to proteins different from those used in training, including proteins for which only a single structure is available. Indeed, when evaluating the performance of our model (Section \ref{sec:test}), we consider only proteins that were not used in training. For many of these proteins, only a single structure is available.

To assign a label to each training input, we first use Glide to calculate the VDW score (i.e., VDW interaction energy) for the candidate pose superimposed on each available holo structure for the given protein. We then determine the minimum value across these scores --- that is, the most favorable score. We use this minimum value as the label (see Figure \ref{supfig:pipeline}).

Formally, we define the minimum VDW score as
\begin{equation}
VDW'(L) = \min_{\forall p_i \in \{p_1, \dots, p_N\}} VDW(L, p_i)
\label{eq:min_vdw}
\end{equation}
where $VDW(L, p_i)$ is the Glide VDW score of a candidate ligand pose $L$ with respect to a target protein structure $p_i$. 

When testing our model --- and when deploying it for drug discovery and biology applications --- we are given only a single structure of the target protein. Because of how the model is trained (on different proteins), however, it effectively predicts what the most favorable VDW score of that pose would be if multiple structures of the target protein were available. In other words, our model implicitly predicts flexibility of a protein's binding pocket given only a single structure of the protein.

\subsection{Datasets} \label{sec:datasets}
Our training, validation, and test datasets consist of sets of poses of ligands docked to protein structures. The protein structures and small molecule ligands used to generate our ligand pose datasets were obtained from the PDBBind 2019 refined dataset \citep{pdbbind}, a collection of protein-ligand complex structures with high resolution. The protein-ligand complex structures are categorized based on the protein (i.e., holo structures of a target protein are grouped together), and those proteins that have at least two holo structures are selected. See Figure \ref{supfig:holo_dist} for distribution of the number of holo structures per unique protein used to generate the labels and ligand docking poses in the training and validation sets. In addition, we also included the benchmark set from \cite{paggi2021leveraging} in our test dataset to ensure good coverage of major drug target protein families: GPCRs, kinases, ion channels and nuclear receptors \citep{santos2017comprehensive}. To ensure no data leakage, we split the proteins for training, validation, and testing such that no protein in the test dataset had more than 30\% sequence identity with any protein in training or validation datasets. There are 228, 85 and 73 unique proteins in the training, validation and test datasets, respectively. 


Next, candidate ligand poses for training and validation are generated using Glide SP \citep{friesner2004glide} with default parameters and then overlaid with a randomly selected holo structure of the same protein to generate poses with and without clashes with the receptor. For each query ligand, a maximum of five protein structures were randomly selected for docking, and 25 poses were randomly selected from each docking result. We follow the procedures described in \cite{paggi2021leveraging} for preparing protein-ligand complex structures and ligands for docking.

Unlike in training/validation, in testing we are given only a single structure of the protein target on which to dock the query ligand. We can only use this one protein structure to generate candidate binding poses for the ligand. Therefore, in addition to (1) generating poses with Glide SP and normal (default) VDW parameters (VDW radius scaling of 1.0/0.8 for receptor/ligand), we ran (2) Glide SP with softened VDW parameters (VDW radius scaling of 0.6/0.5 for receptor/ligand) with extended sampling to generate candidate ligand poses with collisions with the target protein. For each scheme, we set the maximum number of candidate poses to 300 for each protein-ligand pair (referred to here as a "cross-docking pair"). Additionally, we also included a native pose of each query ligand in the candidate pose set, refined with an energy minimization protocol, since otherwise only about 80\% of the cross-docking pairs have any near-native poses among the set of candidate poses generated by the two schemes above (see Figure \ref{supfig:glide_top_n}). On average, we generate roughly 500 candidate ligand binding poses in total for each cross-docking pair.  

For each protein-ligand pair in the test set, we randomly select one protein structure for docking. We ensure that this structure was determined experimentally in the presence of ligand substantially different from the (docked) query ligand --- in particular, that the two ligands have a Tanimoto coefficient of less than 0.4, where the Tanimoto coefficient is computed by comparing the Extended-Connectivity Fingerprints (ECFPs) of the two ligands. This results in 615 cross-docking pairs, which are further divided into two cases: (1) "difficult" cross-docking pairs, defined as those for which the native ligand binding pose, after energy minimization in the docking structure, still exhibits severe clashes with protein atoms (specifically, when the ratio of the distance between two atoms and the sum of their VDW radii is $\leq 0.75$), or where the ligand pose drifts significantly during energy minimization such that it exhibits an RMSD > 2.0\AA\ relative to the original (experimentally determined) ligand pose; (2) "other" cross-docking pairs, defined as the remaining ones. In the "difficult" cases, we expect significant deformation of the protein upon ligand binding, while we expect less protein deformation in the "other" cases.

\subsection{Architecture} \label{sec:architecture}

The input to our ML model is a candidate pose for a ligand and a single protein structure to be used for docking (see Figure \ref{supfig:pipeline}). We also provide our model with the corresponding Glide VDW score. Although we utilize all available holo structures of a target protein to create our training labels (i.e., to calculate VDW', as described in equation \ref{eq:min_vdw}), we do not use these other structures in any way to make the prediction. This reflects the situation in practice, where often only one structure is available to dock the ligand of interest. 

\begin{figure}[h]
\centering
\includegraphics[width=0.75\textwidth]{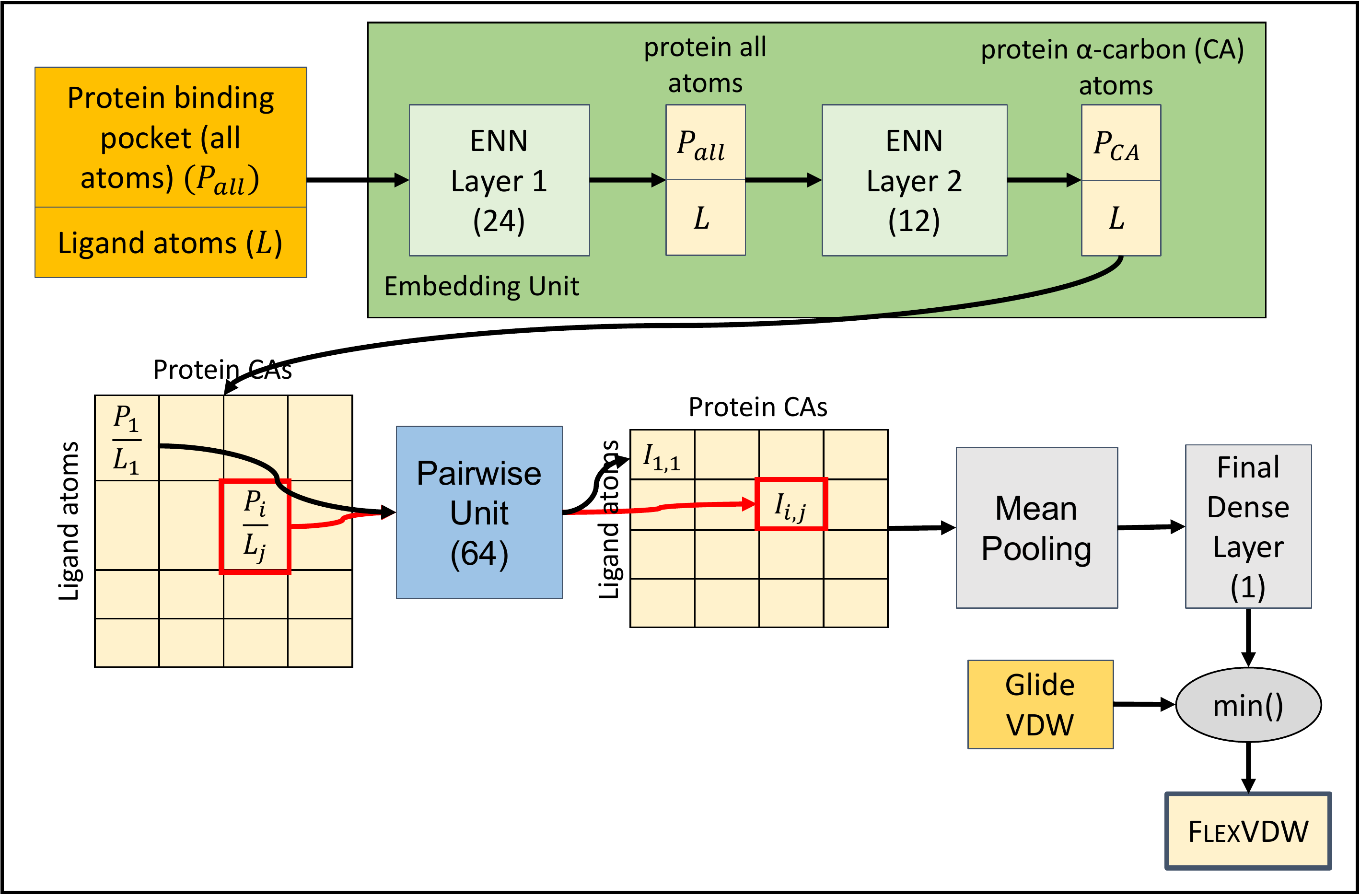}
\caption{Schematic of the architecture of \textsc{FlexVDW} network. The output dimensions of the individual layer are indicated in parentheses. At input, the model takes in a single protein structure and a single candidate ligand pose (orange block) to predict the VDW interaction energy of the ligand pose with respect to the protein structure. The model featurizes the input into basic element type of the atom (C, O, N, P, S, polar H, and F/Cl/Br), secondary structure (if applicable), partial charge and a protein/ligand Boolean flag for each atom. \textsc{FlexVDW} consists of two main components: (1) embedding unit (green block) and (2) pairwise unit (blue block). At the core of the embedding unit are 3D equivariant convolution layers (ENN Layers 1 and 2; light green blocks) that operate on the atomic point cloud to learn the embedding of protein/ligand atoms, which is then used to predict the VDW score of the ligand pose. To regularize our networks, we downsample the protein from all atoms ($P_{all}$) to $\alpha$-carbons ($P_{CA}$) at the last layer of the embedding unit (ENN Layer 2), and apply the same learned function to each protein CA--ligand atom pair (i.e., the pairwise unit) to mimic the pairwise form of physical VDW interactions. In addition, we calculate the Glide VDW score and pass it as input to the \texttt{min()} function in the last layer to make the final prediction. For details on the architecture and the hyperparameters used for each component of the architecture, see Supplement \ref{sup:architecture} and Figure \ref{supfig:architecture-detail}.}
\label{fig:architecture-all}
\end{figure}

Our architecture has two main components: (1) the embedding unit (see Figure \ref{fig:architecture-all}: green block) and (2) the pairwise unit (see Figure \ref{fig:architecture-all}: blue block). The embedding unit learns an embedding of a protein-ligand pose structure, which is then passed to the pairwise unit. At the core of the embedding unit are 3D equivariant convolution layers (ENN Layers 1 and 2) that operate on a 3D atomic point cloud. This point representation in 3D space allows us to accurately represent the relative positioning of atoms in the protein-ligand complex, which is important for capturing the interactions between protein atoms and ligand atoms. Each ENN layer consists of the sequential application of self-interaction, point convolution, point normalization, self-interaction, and nonlinearity \citep{psp}. Each atom/point in 3D is associated with a feature vector. At input, the model takes as features the basic element type of the atom (C, O, N, P, S, polar H, or halogen (F/Cl/Br)) encoded as a one-hot vector, the secondary structure (if applicable), the partial charge of each atom, and a Boolean flag indicating whether the atom belongs to the ligand or the protein. The point-wise feature vectors are updated through the ENN layers by aggregating local information of the nearest 50 neighboring points.

To regularize our networks, we downsample the protein from all atoms to the $\alpha$ carbon (CA) of each amino acid residue in the last ENN layer (ENN Layer 2) of the embedding unit and apply the same learned function to each protein CA--ligand atom pair (i.e., the pairwise unit) to mimic the pairwise form of physical VDW interactions. More concretely, for each protein CA--ligand atom pair, their embeddings from the previous embedding unit are concatenated as input to the pairwise unit, a series of dense neural network layers, to compute their pairwise "interaction" features. These pairwise interaction features are averaged over all pairs (see Figure \ref{fig:architecture-all}: Mean Pooling) and passed through the final dense neural network layer (see Figure \ref{fig:architecture-all}: Final Dense Layer) to obtain a single scalar prediction. Inspired by \cite{wang2019machine,husic2020coarse}, which use a prior energy for learning molecular dynamics force fields, we use additional information from the Glide VDW score and pass it as input to the \texttt{min()} function in the last layer along with the output from the Final Dense Layer in order to make the final prediction.

For details on the architecture and the hyperparameters used for each component of the architecture, see Supplement \ref{sup:architecture} and Figure \ref{supfig:architecture-detail}

\subsection{Training}

We formulate the training as a regression task aimed at predicting VDW', the minimum of the candidate ligand's VDW score over several available holo structures of the protein. The MSE loss between the actual and predicted values of VDW' is used as a loss function. To prevent loss explosion during training, the training label is capped at 100; otherwise, it could occasionally be on the order of a million or more. We train with the Adam optimizer in PyTorch (\cite{pytorch}) with a learning rate of 0.00005 and a batch size of 4 for 10 epochs and monitor the loss on the validation set at every epoch. In the first 5 epochs, the input Glide VDW score is ignored, in order to prevent the model from overfitting to the Glide VDW score instead of learning about protein flexibility. In the next 5 epochs, the Glide VDW score is added. The weights of the network that performs best on the validation set are then used to evaluate the predictions on the test set. We train the models on one NVIDIA GeForce RTX 3090 GPU for around 20 hours.

\section{Results}

\subsection{Evaluation of cross-docking results on test set} \label{sec:test}

To evaluate the strength of our machine-learned scoring function, \textsc{FlexVDW}, in terms of docking accuracy, we evaluate the top-N near-native hit rate, which is defined as the fraction of cross-docking cases for which a near-native pose is included in the first N poses when the poses are ranked by the docking score. Here, we consider a pose to be near native if its root mean square deviation (RMSD) from the experimentally determined pose is less than or equal to 2.0\AA\ (a threshold commonly used in practice \citep{kontoyianni2004evaluation,cole2005comparing}).

\begin{figure}[h]
\centering
\includegraphics[width=1\textwidth]{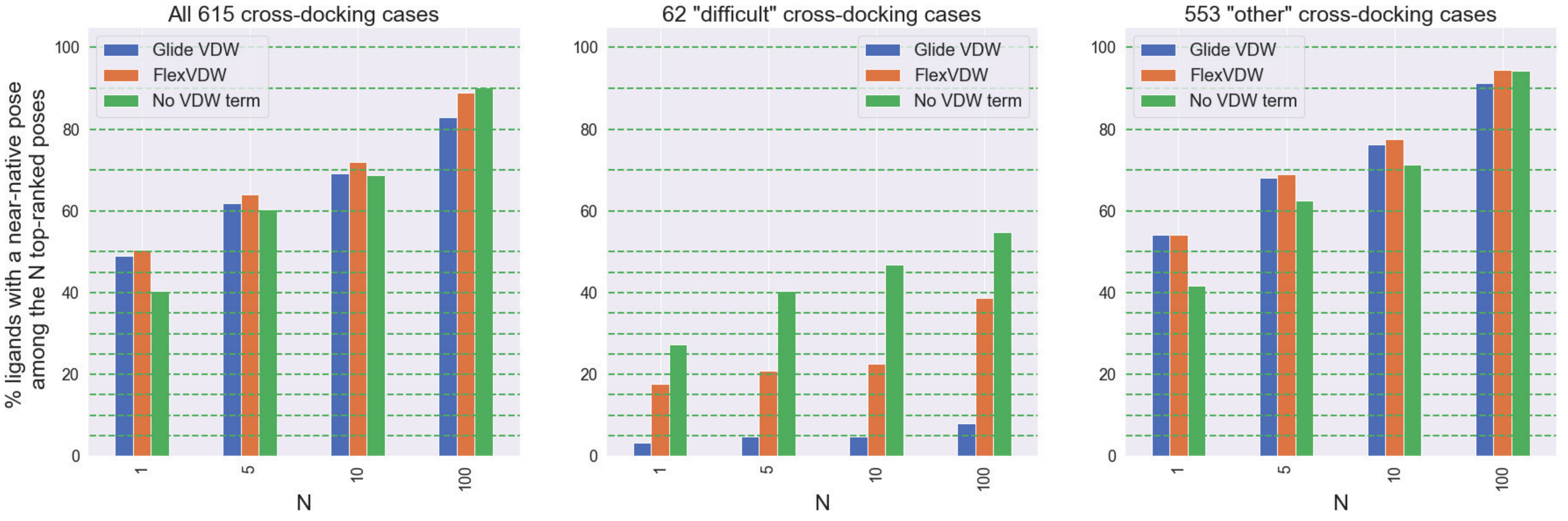}
\caption{Percentage of cases for which a near-native pose is included in the top N poses sorted by docking score (higher is better). Our approach significantly improves over Glide performance in "difficult" cross-docking cases where significant deformation of the protein is expected upon ligand binding, while maintaining performance in "other" cross-docking cases where minimal deformation of the protein is expected. Although the absence of a VDW term in the docking score leads to better performance in "difficult" cases, it worsens overall performance (especially for the top-ranked pose), showing the importance of including a VDW term in the docking score.}
\label{fig:ml_soft2}
\end{figure}

The evaluation is performed on the candidate ligand poses generated for the 615 cross-docking pairs in the test dataset (see Section \ref{sec:datasets}). During testing, only a single protein structure is provided to our ML model. We compare performance of the Glide scoring function with its original VDW term and with that term replaced by \textsc{FlexVDW}. As can be seen in Figure \ref{fig:ml_soft2},  incorporation of \textsc{FlexVDW} into Glide improves performance in "difficult" cross-docking cases (middle panel), where significant deformation of the protein is typically required upon ligand binding. At the same time, \textsc{FlexVDW} achieves performance similar to that of Glide's original VDW term for the "other" cross-docking cases where less protein deformation is typically required (right panel).

In addition, as a baseline, we evaluate the accuracy of a scoring function in which we simply remove the VDW term while keeping the other terms of the Glide scoring function. As we can see in Figure \ref{fig:ml_soft2}, although eliminating the VDW term leads to a better top-N near-native hit rate for "difficult" cases compared to \textsc{FlexVDW}, overall performance deteriorates (especially for the top-1 near-native hit rate), which shows the importance of including a VDW term in the docking score. As we allow more ligand poses with severe collisions with protein backbones ("garbage poses") in the candidate pose set, the performance of \textsc{FlexVDW} decreases, but the performance of the docking score without a VDW term decreases even more, showing that our approach is able to generalize to some extent even if we never train the model with "garbage" poses, and further highlighting the importance of the VDW term in the docking score (see Figure \ref{supfig:ml_softest}).




\subsection{Comparison of Glide and \textsc{FlexVDW} predicted scores and top-1 poses}

\begin{figure}[h]
\centering
\includegraphics[width=0.65\textwidth]{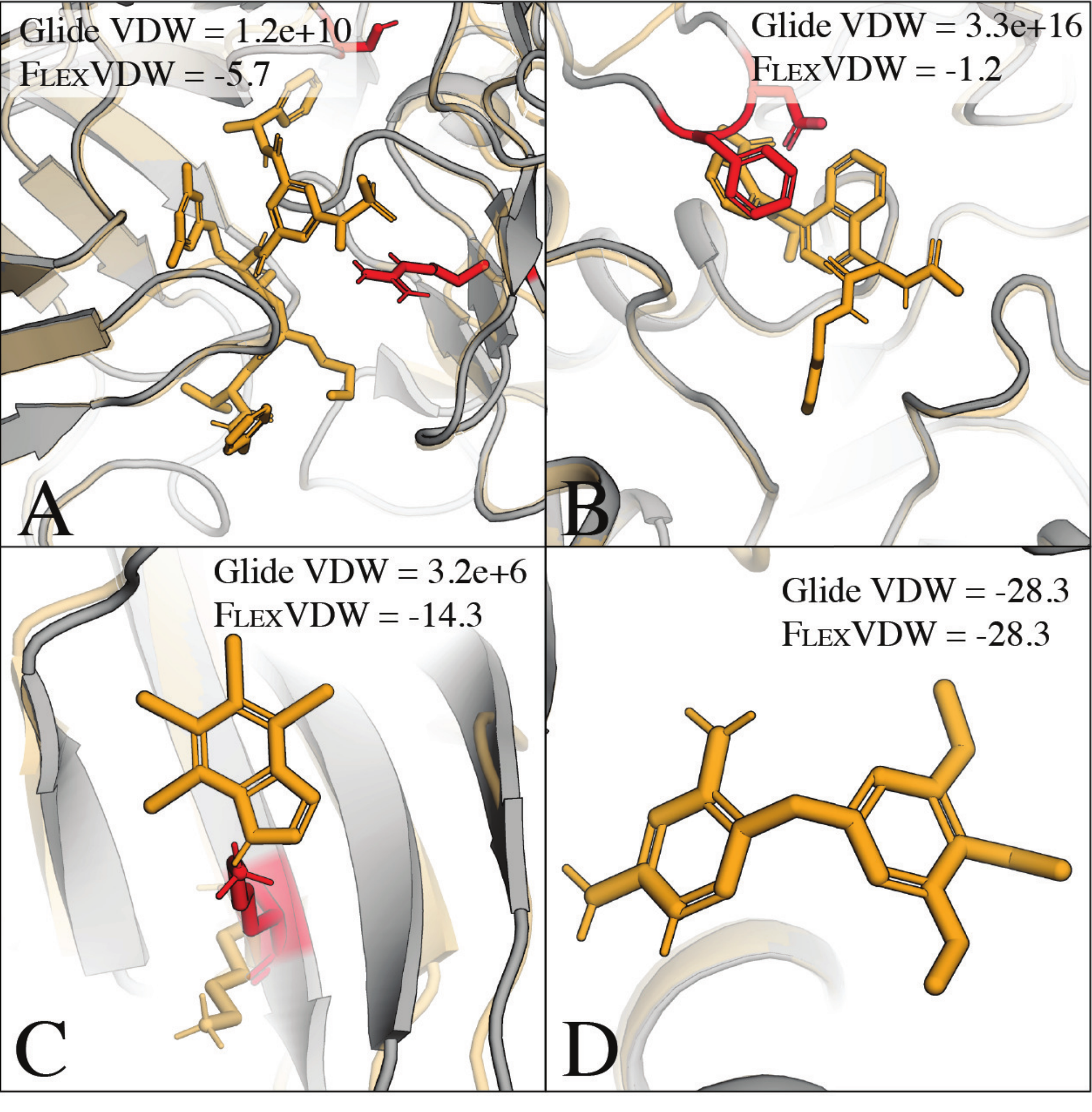}
\caption{Comparison of the Glide VDW and \textsc{FlexVDW} score of the native ligand poses (orange) when docked to other holo structures of the same protein (grey). The protein side chains that clash with the native ligand poses are shown in red. (A): Native ligand pose of 3I25 docked to the protein structure of 1FKN (Uniprot ID: P56817). The native ligand pose clashes with several residue side chains. (B): Native ligand pose of 1NL9 docked to the protein structure of 2F6T (Uniprot ID: P18031). The native ligand pose clashes with the loop that should have been located further up in the crystal structure of 2F6T (transparent light orange). (C): Native ligand pose of 2OXY docked to the protein structure of 2PVJ (Uniprot ID: P28523). The native ligand pose clashes with the lysine side chain (red) in the docking structure. In the native crystal structure, the lysine side chain points downward (transparent light orange). (D): Native ligand pose of 2W3A docked to the protein structure of 1BOZ (Uniprot ID: P00374). Here, there is no clash between the native ligand pose and the protein. Both Glide and \textsc{FlexVDW} give the same VDW scores of -28.3.}
\label{fig:ml_vs_glide_vdw}
\end{figure}

Next, we compare the \textsc{FlexVDW} and Glide VDW scores for the native ligand poses when superimposed on structures of the target protein determined with other ligands bound. In Figure \ref{fig:ml_vs_glide_vdw}A-C, the native ligand poses clash with the docking structures. Glide assigns very high (unfavorable) VDW scores, preventing it from predicting these poses. Indeed, in these cases, Glide's top-ranked (top-1) ligand pose predictions differ substantially from the native pose(see Figure~ \ref{supfig:ml_vs_glide_top_1}A-C). In contrast, our machine-learned predictor, \textsc{FlexVDW}, handles these cases better. In two of the three cases, it ranks near-native poses first (top-1) (see Figure \ref{supfig:ml_vs_glide_top_1}A and C). In the third case (Figure \ref{supfig:ml_vs_glide_top_1}B), even though \textsc{FlexVDW} predicts a negative VDW score for the native ligand pose (see Figure \ref{fig:ml_vs_glide_vdw}B), the near-native poses are eventually rejected due to the high electrostatic repulsion energy, thus \textsc{FlexVDW} fails to select the near-native pose as the top-1 pose. When there is no clash between the ligand pose and the protein structure used for docking, \textsc{FlexVDW} is comparable to Glide in ranking near-native ligand pose highly (see Figure \ref{fig:ml_vs_glide_vdw}D and ~\ref{supfig:ml_vs_glide_top_1}D).

\section{Discussion}

We have demonstrated the feasibility of learning a scoring function that accounts for protein flexibility in ligand docking without explicitly modeling changes in protein structure. Given a protein structure and a candidate ligand pose, \textsc{FlexVDW} predicts the VDW value of this ligand pose, taking into account the flexibility of the protein. 

To evaluate the strength of our machine-learned scoring function in terms of docking accuracy, we evaluate the top-N near-native hit rate of cross-docking protein-ligand pairs. To ensure the generalizability of our methods to different protein families, we select our test cases to cover the major drug target protein families, including GPCRs, kinases, ion channels, nuclear receptors, and others. We show that incorporating this machine-learned VDW term into Glide, a state-of-the-art physics-based scoring function, improves docking accuracy in cases with substantial protein deformation upon ligand binding, without degrading performance in cases with minimal protein deformation upon ligand binding. Our approach could be integrated with any existing physics-based scoring function, not limited to Glide, with refitting to the particular physics-based scoring function of interest. 

There are several limitations to our approach. First, we formulate our learning task in terms of predicting the global VDW score. Reformulating the learning task in terms of predicting VDW interaction energies between individual pairs of atoms could potentially provide a better signal for which parts of the protein are flexible upon ligand binding. Additionally, we consider only VDW interactions and ignore the electrostatic interaction. In some cases, a near-native ligand pose is eliminated not only due to a high VDW energy, but also due to a high electrostatic repulsion energy. Future work is necessary to address these issues.

Second, because we assign training labels using the minimum VDW score across multiple holo structures as a proxy for protein flexibility, the extent to which our model can learn about protein flexibility is limited by the diversity of available holo structures. This could potentially be improved by including snapshots from molecular dynamics simulations as additional protein structures when determining the training labels. 

Note that when using our model to predict ligand binding poses, we use only a single structure of the target protein --- because often only one a single structure of a given protein is available. Indeed, when when evaluating the performance of our model, we use only a single  structure for each protein, and the proteins used for evaluation are all substantially different from those used to train the model.

In summary, our work is a step toward incorporating implicit protein flexibility into ligand docking, which will improve the accuracy of ligand binding pose prediction.


\subsubsection*{Funding Information}
PS was supported by a Graduate Research Fellowship from the US National Science Foundation (NSF). JMP was supported by a Stanford Graduate Fellowship.

\bibliography{iclr2023_conference}

\begin{thebibliography}{38}
\providecommand{\natexlab}[1]{#1}
\providecommand{\url}[1]{\texttt{#1}}
\expandafter\ifx\csname urlstyle\endcsname\relax
  \providecommand{\doi}[1]{doi: #1}\else
  \providecommand{\doi}{doi: \begingroup \urlstyle{rm}\Url}\fi

\bibitem[Allen et~al.(2015)Allen, Balius, Mukherjee, Brozell, Moustakas, Lang,
  Case, Kuntz, and Rizzo]{allen2015dock}
William~J Allen, Trent~E Balius, Sudipto Mukherjee, Scott~R Brozell, Demetri~T
  Moustakas, P~Therese Lang, David~A Case, Irwin~D Kuntz, and Robert~C Rizzo.
\newblock Dock 6: Impact of new features and current docking performance.
\newblock \emph{Journal of computational chemistry}, 36\penalty0 (15):\penalty0
  1132--1156, 2015.

\bibitem[Amaro et~al.(2018)Amaro, Baudry, Chodera, Demir, McCammon, Miao, and
  Smith]{amaro2018ensemble}
Rommie~E Amaro, Jerome Baudry, John Chodera, {\"O}zlem Demir, J~Andrew
  McCammon, Yinglong Miao, and Jeremy~C Smith.
\newblock Ensemble docking in drug discovery.
\newblock \emph{Biophysical journal}, 114\penalty0 (10):\penalty0 2271--2278,
  2018.

\bibitem[Bender et~al.(2021)Bender, Gahbauer, Luttens, Lyu, Webb, Stein, Fink,
  Balius, Carlsson, Irwin, et~al.]{bender2021practical}
Brian~J Bender, Stefan Gahbauer, Andreas Luttens, Jiankun Lyu, Chase~M Webb,
  Reed~M Stein, Elissa~A Fink, Trent~E Balius, Jens Carlsson, John~J Irwin,
  et~al.
\newblock A practical guide to large-scale docking.
\newblock \emph{Nature protocols}, 16\penalty0 (10):\penalty0 4799--4832, 2021.

\bibitem[Clevert et~al.(2016)Clevert, Unterthiner, and Hochreiter]{elu}
Djork-Arn{\'e} Clevert, Thomas Unterthiner, and Sepp Hochreiter.
\newblock Fast and accurate deep network learning by exponential linear units
  (elus).
\newblock \emph{arXiv: Learning}, 2016.

\bibitem[Cole et~al.(2005)Cole, Murray, Nissink, Taylor, and
  Taylor]{cole2005comparing}
Jason~C Cole, Christopher~W Murray, J~Willem~M Nissink, Richard~D Taylor, and
  Robin Taylor.
\newblock Comparing protein--ligand docking programs is difficult.
\newblock \emph{Proteins: Structure, Function, and Bioinformatics}, 60\penalty0
  (3):\penalty0 325--332, 2005.

\bibitem[Coleman et~al.(2013)Coleman, Carchia, Sterling, Irwin, and
  Shoichet]{coleman2013ligand}
Ryan~G Coleman, Michael Carchia, Teague Sterling, John~J Irwin, and Brian~K
  Shoichet.
\newblock Ligand pose and orientational sampling in molecular docking.
\newblock \emph{PloS one}, 8\penalty0 (10):\penalty0 e75992, 2013.

\bibitem[Eismann et~al.(2020)Eismann, Suriana, Jing, Townshend, and Dror]{psp}
Stephan Eismann, Patricia Suriana, Bowen Jing, Raphael J.~L. Townshend, and
  Ron~O. Dror.
\newblock Protein model quality assessment using rotation-equivariant,
  hierarchical neural networks.
\newblock \emph{arXiv preprint arXiv:2011.13557}, 2020.

\bibitem[Eismann et~al.(2021)Eismann, Townshend, Thomas, Jagota, Jing, and
  Dror]{eismann2020hierarchical}
Stephan Eismann, Raphael~J.L. Townshend, Nathaniel Thomas, Milind Jagota, Bowen
  Jing, and Ron~O. Dror.
\newblock Hierarchical, rotation-equivariant neural networks to select
  structural models of protein complexes.
\newblock \emph{Proteins: Structure, Function, and Bioinformatics}, 89\penalty0
  (5):\penalty0 493--501, 2021.
\newblock \doi{https://doi.org/10.1002/prot.26033}.
\newblock URL \url{https://onlinelibrary.wiley.com/doi/abs/10.1002/prot.26033}.

\bibitem[Elokely \& Doerksen(2013)Elokely and Doerksen]{elokely2013docking}
Khaled~M Elokely and Robert~J Doerksen.
\newblock Docking challenge: protein sampling and molecular docking
  performance.
\newblock \emph{Journal of chemical information and modeling}, 53\penalty0
  (8):\penalty0 1934--1945, 2013.

\bibitem[Evangelista~Falcon et~al.(2019)Evangelista~Falcon, Ellingson, Smith,
  and Baudry]{evangelista2019ensemble}
Wilfredo Evangelista~Falcon, Sally~R Ellingson, Jeremy~C Smith, and Jerome
  Baudry.
\newblock Ensemble docking in drug discovery: how many protein configurations
  from molecular dynamics simulations are needed to reproduce known ligand
  binding?
\newblock \emph{The Journal of Physical Chemistry B}, 123\penalty0
  (25):\penalty0 5189--5195, 2019.

\bibitem[Ferreira et~al.(2015)Ferreira, Dos~Santos, Oliva, and
  Andricopulo]{ferreira2015molecular}
Leonardo~G Ferreira, Ricardo~N Dos~Santos, Glaucius Oliva, and Adriano~D
  Andricopulo.
\newblock Molecular docking and structure-based drug design strategies.
\newblock \emph{Molecules}, 20\penalty0 (7):\penalty0 13384--13421, 2015.

\bibitem[Forli et~al.(2016)Forli, Huey, Pique, Sanner, Goodsell, and
  Olson]{forli2016computational}
Stefano Forli, Ruth Huey, Michael~E Pique, Michel~F Sanner, David~S Goodsell,
  and Arthur~J Olson.
\newblock Computational protein--ligand docking and virtual drug screening with
  the autodock suite.
\newblock \emph{Nature protocols}, 11\penalty0 (5):\penalty0 905--919, 2016.

\bibitem[Friesner et~al.(2004)Friesner, Banks, Murphy, Halgren, Klicic, Mainz,
  Repasky, Knoll, Shelley, Perry, et~al.]{friesner2004glide}
Richard~A Friesner, Jay~L Banks, Robert~B Murphy, Thomas~A Halgren, Jasna~J
  Klicic, Daniel~T Mainz, Matthew~P Repasky, Eric~H Knoll, Mee Shelley, Jason~K
  Perry, et~al.
\newblock Glide: a new approach for rapid, accurate docking and scoring. 1.
  method and assessment of docking accuracy.
\newblock \emph{Journal of medicinal chemistry}, 47\penalty0 (7):\penalty0
  1739--1749, 2004.

\bibitem[Guedes et~al.(2014)Guedes, de~Magalh{\~a}es, and
  Dardenne]{guedes2014receptor}
Isabella~A Guedes, Camila~S de~Magalh{\~a}es, and Laurent~E Dardenne.
\newblock Receptor--ligand molecular docking.
\newblock \emph{Biophysical reviews}, 6:\penalty0 75--87, 2014.

\bibitem[Husic et~al.(2020)Husic, Charron, Lemm, Wang, P{\'e}rez, Majewski,
  Kr{\"a}mer, Chen, Olsson, de~Fabritiis, et~al.]{husic2020coarse}
Brooke~E Husic, Nicholas~E Charron, Dominik Lemm, Jiang Wang, Adri{\`a}
  P{\'e}rez, Maciej Majewski, Andreas Kr{\"a}mer, Yaoyi Chen, Simon Olsson,
  Gianni de~Fabritiis, et~al.
\newblock Coarse graining molecular dynamics with graph neural networks.
\newblock \emph{The Journal of chemical physics}, 153\penalty0 (19):\penalty0
  194101, 2020.

\bibitem[Jones et~al.(1997)Jones, Willett, Glen, Leach, and
  Taylor]{jones1997development}
Gareth Jones, Peter Willett, Robert~C Glen, Andrew~R Leach, and Robin Taylor.
\newblock Development and validation of a genetic algorithm for flexible
  docking.
\newblock \emph{Journal of molecular biology}, 267\penalty0 (3):\penalty0
  727--748, 1997.

\bibitem[Jordan et~al.(2016)Jordan, Whittington, Bartberger, Sickmier, Chen,
  Cheng, and Judd]{jordan2016fragment}
John~B Jordan, Douglas~A Whittington, Michael~D Bartberger, E~Allen Sickmier,
  Kui Chen, Yuan Cheng, and Ted Judd.
\newblock Fragment-linking approach using 19f nmr spectroscopy to obtain highly
  potent and selective inhibitors of $\beta$-secretase.
\newblock \emph{Journal of Medicinal Chemistry}, 59\penalty0 (8):\penalty0
  3732--3749, 2016.

\bibitem[Khamis et~al.(2015)Khamis, Gomaa, and Ahmed]{khamis2015machine}
Mohamed~A Khamis, Walid Gomaa, and Walaa~F Ahmed.
\newblock Machine learning in computational docking.
\newblock \emph{Artificial intelligence in medicine}, 63\penalty0 (3):\penalty0
  135--152, 2015.

\bibitem[Kitchen et~al.(2004)Kitchen, Decornez, Furr, and
  Bajorath]{kitchen2004docking}
Douglas~B Kitchen, H{\'e}l{\`e}ne Decornez, John~R Furr, and J{\"u}rgen
  Bajorath.
\newblock Docking and scoring in virtual screening for drug discovery: methods
  and applications.
\newblock \emph{Nature reviews Drug discovery}, 3\penalty0 (11):\penalty0
  935--949, 2004.

\bibitem[Kontoyianni et~al.(2004)Kontoyianni, McClellan, and
  Sokol]{kontoyianni2004evaluation}
Maria Kontoyianni, Laura~M McClellan, and Glenn~S Sokol.
\newblock Evaluation of docking performance: comparative data on docking
  algorithms.
\newblock \emph{Journal of medicinal chemistry}, 47\penalty0 (3):\penalty0
  558--565, 2004.

\bibitem[Korb et~al.(2012)Korb, Olsson, Bowden, Hall, Verdonk, Liebeschuetz,
  and Cole]{korb2012potential}
Oliver Korb, Tjelvar~SG Olsson, Simon~J Bowden, Richard~J Hall, Marcel~L
  Verdonk, John~W Liebeschuetz, and Jason~C Cole.
\newblock Potential and limitations of ensemble docking.
\newblock \emph{Journal of chemical information and modeling}, 52\penalty0
  (5):\penalty0 1262--1274, 2012.

\bibitem[Lemmon \& Meiler(2012)Lemmon and Meiler]{lemmon2012rosetta}
Gordon Lemmon and Jens Meiler.
\newblock Rosetta ligand docking with flexible xml protocols.
\newblock \emph{Computational Drug Discovery and Design}, pp.\  143--155, 2012.

\bibitem[Liu et~al.(2015)Liu, Li, Han, Li, Liu, Zhao, Nie, Liu, and
  Wang]{pdbbind}
Zhihai Liu, Yan Li, Li~Han, Jie Li, Jie Liu, Zhixiong Zhao, Wei Nie, Yuchen
  Liu, and Renxiao Wang.
\newblock {PDB-wide} collection of binding data: current status of the
  {PDBbind} database.
\newblock \emph{Bioinformatics}, 31\penalty0 (3):\penalty0 405--412, February
  2015.

\bibitem[McNutt et~al.(2021)McNutt, Francoeur, Aggarwal, Masuda, Meli, Ragoza,
  Sunseri, and Koes]{mcnutt2021gnina}
Andrew~T McNutt, Paul Francoeur, Rishal Aggarwal, Tomohide Masuda, Rocco Meli,
  Matthew Ragoza, Jocelyn Sunseri, and David~Ryan Koes.
\newblock Gnina 1.0: molecular docking with deep learning.
\newblock \emph{Journal of cheminformatics}, 13\penalty0 (1):\penalty0 1--20,
  2021.

\bibitem[Miller et~al.(2021)Miller, Murphy, Sindhikara, Borrelli, Grisewood,
  Ranalli, Dixon, Jerome, Boyles, Day, et~al.]{miller2021reliable}
Edward~B Miller, Robert~B Murphy, Daniel Sindhikara, Kenneth~W Borrelli,
  Matthew~J Grisewood, Fabio Ranalli, Steven~L Dixon, Steven Jerome, Nicholas~A
  Boyles, Tyler Day, et~al.
\newblock Reliable and accurate solution to the induced fit docking problem for
  protein--ligand binding.
\newblock \emph{Journal of Chemical Theory and Computation}, 17\penalty0
  (4):\penalty0 2630--2639, 2021.

\bibitem[Morrone et~al.(2020)Morrone, Weber, Huynh, Luo, and
  Cornell]{morrone2020combining}
Joseph~A Morrone, Jeffrey~K Weber, Tien Huynh, Heng Luo, and Wendy~D Cornell.
\newblock Combining docking pose rank and structure with deep learning improves
  protein--ligand binding mode prediction over a baseline docking approach.
\newblock \emph{Journal of chemical information and modeling}, 60\penalty0
  (9):\penalty0 4170--4179, 2020.

\bibitem[Novoa et~al.(2010)Novoa, Pouplana, Barril, and
  Orozco]{novoa2010ensemble}
Eva~Maria Novoa, Lluis Ribas~de Pouplana, Xavier Barril, and Modesto Orozco.
\newblock Ensemble docking from homology models.
\newblock \emph{Journal of Chemical Theory and Computation}, 6\penalty0
  (8):\penalty0 2547--2557, 2010.

\bibitem[Paggi et~al.(2021)Paggi, Belk, Hollingsworth, Villanueva, Powers,
  Clark, Chemparathy, Tynan, Lau, Sunahara, et~al.]{paggi2021leveraging}
Joseph~M Paggi, Julia~A Belk, Scott~A Hollingsworth, Nicolas Villanueva,
  Alexander~S Powers, Mary~J Clark, Augustine~G Chemparathy, Jonathan~E Tynan,
  Thomas~K Lau, Roger~K Sunahara, et~al.
\newblock Leveraging nonstructural data to predict structures and affinities of
  protein--ligand complexes.
\newblock \emph{Proceedings of the National Academy of Sciences}, 118\penalty0
  (51):\penalty0 e2112621118, 2021.

\bibitem[Park et~al.(2008)Park, Min, Kwak, Koo, Lim, Seo, Choi, Platt, and
  Choi]{park2008synthesis}
Heuisul Park, Kyeongsik Min, Hyo-Shin Kwak, Ki~Dong Koo, Dongchul Lim, Sang-Won
  Seo, Jae-Ung Choi, Bettina Platt, and Deog-Young Choi.
\newblock Synthesis, sar, and x-ray structure of human bace-1 inhibitors with
  cyclic urea derivatives.
\newblock \emph{Bioorganic \& medicinal chemistry letters}, 18\penalty0
  (9):\penalty0 2900--2904, 2008.

\bibitem[Paszke et~al.(2019)Paszke, Gross, Massa, Lerer, Bradbury, Chanan,
  Killeen, Lin, Gimelshein, Antiga, Desmaison, Kopf, Yang, DeVito, Raison,
  Tejani, Chilamkurthy, Steiner, Fang, Bai, and Chintala]{pytorch}
Adam Paszke, Sam Gross, Francisco Massa, Adam Lerer, James Bradbury, Gregory
  Chanan, Trevor Killeen, Zeming Lin, Natalia Gimelshein, Luca Antiga, Alban
  Desmaison, Andreas Kopf, Edward Yang, Zachary DeVito, Martin Raison, Alykhan
  Tejani, Sasank Chilamkurthy, Benoit Steiner, Lu~Fang, Junjie Bai, and Soumith
  Chintala.
\newblock Pytorch: An imperative style, high-performance deep learning library.
\newblock In H.~Wallach, H.~Larochelle, A.~Beygelzimer, F.~d\textquotesingle
  Alch\'{e}-Buc, E.~Fox, and R.~Garnett (eds.), \emph{Advances in Neural
  Information Processing Systems 32}, pp.\  8024--8035. Curran Associates,
  Inc., 2019.
\newblock URL
  \url{http://papers.neurips.cc/paper/9015-pytorch-an-imperative-style-high-performance-deep-learning-library.pdf}.

\bibitem[Ragoza et~al.(2017)Ragoza, Hochuli, Idrobo, Sunseri, and
  Koes]{ragoza2017protein}
Matthew Ragoza, Joshua Hochuli, Elisa Idrobo, Jocelyn Sunseri, and David~Ryan
  Koes.
\newblock Protein--ligand scoring with convolutional neural networks.
\newblock \emph{Journal of chemical information and modeling}, 57\penalty0
  (4):\penalty0 942--957, 2017.

\bibitem[Ravindranath et~al.(2015)Ravindranath, Forli, Goodsell, Olson, and
  Sanner]{ravindranath2015autodockfr}
Pradeep~Anand Ravindranath, Stefano Forli, David~S Goodsell, Arthur~J Olson,
  and Michel~F Sanner.
\newblock Autodockfr: advances in protein-ligand docking with explicitly
  specified binding site flexibility.
\newblock \emph{PLoS computational biology}, 11\penalty0 (12):\penalty0
  e1004586, 2015.

\bibitem[Santos et~al.(2017)Santos, Ursu, Gaulton, Bento, Donadi, Bologa,
  Karlsson, Al-Lazikani, Hersey, Oprea, et~al.]{santos2017comprehensive}
Rita Santos, Oleg Ursu, Anna Gaulton, A~Patr{\'\i}cia Bento, Ramesh~S Donadi,
  Cristian~G Bologa, Anneli Karlsson, Bissan Al-Lazikani, Anne Hersey, Tudor~I
  Oprea, et~al.
\newblock A comprehensive map of molecular drug targets.
\newblock \emph{Nature reviews Drug discovery}, 16\penalty0 (1):\penalty0
  19--34, 2017.

\bibitem[Shen et~al.(2020)Shen, Ding, Wang, Cao, Ding, and
  Hou]{shen2020machine}
Chao Shen, Junjie Ding, Zhe Wang, Dongsheng Cao, Xiaoqin Ding, and Tingjun Hou.
\newblock From machine learning to deep learning: Advances in scoring functions
  for protein--ligand docking.
\newblock \emph{Wiley Interdisciplinary Reviews: Computational Molecular
  Science}, 10\penalty0 (1):\penalty0 e1429, 2020.

\bibitem[Totrov \& Abagyan(2008)Totrov and Abagyan]{totrov2008flexible}
Maxim Totrov and Ruben Abagyan.
\newblock Flexible ligand docking to multiple receptor conformations: a
  practical alternative.
\newblock \emph{Current opinion in structural biology}, 18\penalty0
  (2):\penalty0 178--184, 2008.

\bibitem[Trott \& Olson(2010)Trott and Olson]{trott2010autodock}
Oleg Trott and Arthur~J Olson.
\newblock Autodock vina: improving the speed and accuracy of docking with a new
  scoring function, efficient optimization, and multithreading.
\newblock \emph{Journal of computational chemistry}, 31\penalty0 (2):\penalty0
  455--461, 2010.

\bibitem[Verdonk et~al.(2003)Verdonk, Cole, Hartshorn, Murray, and
  Taylor]{verdonk2003improved}
Marcel~L Verdonk, Jason~C Cole, Michael~J Hartshorn, Christopher~W Murray, and
  Richard~D Taylor.
\newblock Improved protein--ligand docking using gold.
\newblock \emph{Proteins: Structure, Function, and Bioinformatics}, 52\penalty0
  (4):\penalty0 609--623, 2003.

\bibitem[Wang et~al.(2019)Wang, Olsson, Wehmeyer, P{\'e}rez, Charron,
  De~Fabritiis, No{\'e}, and Clementi]{wang2019machine}
Jiang Wang, Simon Olsson, Christoph Wehmeyer, Adri{\`a} P{\'e}rez, Nicholas~E
  Charron, Gianni De~Fabritiis, Frank No{\'e}, and Cecilia Clementi.
\newblock Machine learning of coarse-grained molecular dynamics force fields.
\newblock \emph{ACS central science}, 5\penalty0 (5):\penalty0 755--767, 2019.

\end{thebibliography}
\bibliographystyle{iclr2023_conference}

\setcounter{section}{0}
\renewcommand{\thesection}{S\arabic{section}}
\renewcommand{\theHsection}{Supplement.\thesection}

\setcounter{figure}{0}
\renewcommand\thefigure{S\arabic{figure}}
\renewcommand{\theHfigure}{Supplement.\thefigure}

\clearpage
{\Huge{Supplementary Information}}

\section{Details on the Architecture} \label{sup:architecture}

\begin{figure}[h]
\centering
\includegraphics[width=0.88\textwidth]{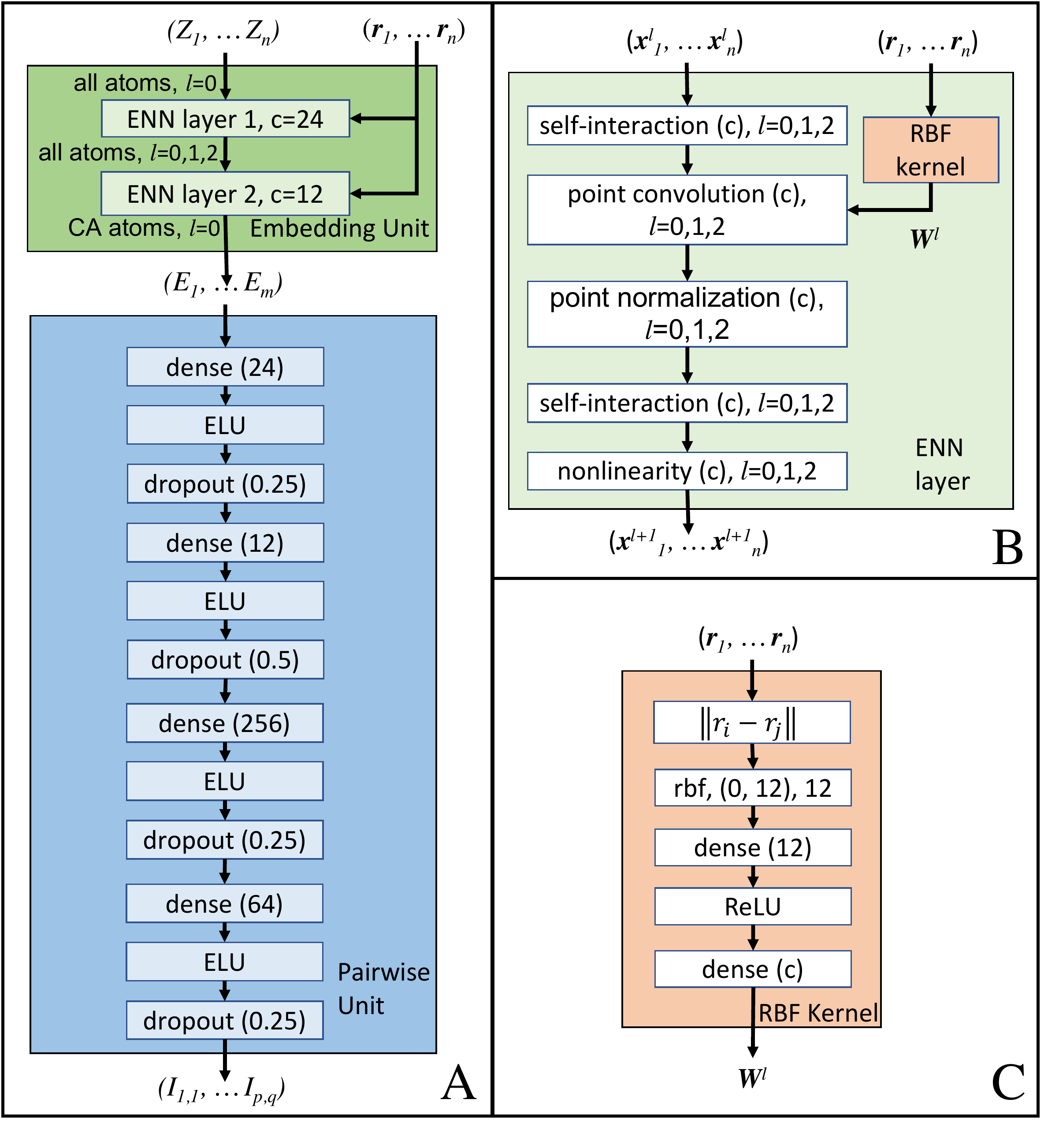}
\caption{Detail Schematic of the architecture of \textsc{FlexVDW} network. The output size of each layer and the dropout parameters are indicated in parentheses. Panel A shows the overall architecture of \textsc{FlexVDW}, which consists of (1) an embedding unit (green block) and (2) a pairwise unit (blue block). We represent a protein-ligand complex as a set of points in 3D with associated scalar features ($Z1, \ldots Z_{n}$). The embedding unit learns an embedding feature for each point through two layers of ENN layers (panel B), each with output channels of size 24 and 12, respectively. For each protein--ligand atom pair, we concatenate their embeddings and pass it to the pairwise unit (blue block) to calculate the pairwise protein--ligand atom interaction features ($I_{1,1}, \ldots I_{p,q}$). These pairwise interaction features are then averaged over all pairs (see Figure \ref{fig:architecture-all}: Mean Pooling) and passed through the Final Dense Layer (see Figure \ref{fig:architecture-all}) to obtain a single scalar prediction. Panel B shows the architecture of the ENN layer. The ENN layer updates the features associated with each point with respect to the features of the 50 nearest neighbor points in 3D Euclidean space, weighted by the distances of the neighbors to the point ($W^l$). Panel C shows the Gaussian radial basis function (RBF) kernel, a trainable network consisting of a hidden dense layer of size 12, a ReLU nonlinearity, and a final dense layer. This kernel computes the weights based on the distances of each point for updating the point convolution (panel B). The number of basis and maximum radius of the Gaussian RBF kernel determines the spatial resolution of the kernel, which we chose to be 12 and 12.0 Å respectively.}
\label{supfig:architecture-detail}
\end{figure}

The ENN layers of the embedding unit (see Figure \ref{supfig:architecture-detail}A: green block) consists of sequential application of  self-interaction, point-convolution, point normalizationm, self-interaction, and nonlinearity \citep{eismann2020hierarchical} (see Figure \ref{supfig:architecture-detail}B). We restrict the maximum filter rotation order at each ENN layer to $l=2$. At each point convolution, it updates the features associated to a given point $p$ based on the features of 50 closest neighboring points in 3D Euclidian space, weighted by their distances to $p$. We express these weighted distances ($W^l$) in terms of Gaussian radial basis function (RBF) kernel, as a trainable network of two dense layers with a hidden layer (followed by a ReLU nonlinearity) of size 12. The number of basis and maximum radius of the Gaussian RBF kernel determines the spatial resolution of the kernel, which we chose to be 12 and 12.0Å respectively for our architecture (see Figure \ref{supfig:architecture-detail}C). 

Starting from a one-hot encoding of the element type,  secondary structure, partial charges (from the OPLS force field), and Boolean protein/ligand flags as feature channels at input to the embedding unit ($Z_1,\ldots Z_{n}$), the first ENN layer of the embedding unit mixes those features and outputs 24 feature channels per rotation order ($l=0$, $l=1$, and $l=2$). The second ENN layer further mixes those features to output 12 feature channels per rotation order (see Figure \ref{supfig:architecture-detail}A: green block). 

Next, we constructed pairwise features of protein-CA--ligand-atom pairs by concatenating their 0-th rotation order ($l=0$) embeddings ($E_1,\ldots E_{n}$), which are then passed to the pairwise unit to calculate the pairwise protein--ligand atom interaction features ($I_{1,1},\ldots I_{p,q}$). This pairwise unit consists of a series of dense neural network layers (followed by the ELU activation function \citep{elu} and a dropout layer). For more details on the layers (including the output feature dimensions of each layer), see Figure \ref{supfig:architecture-detail}A: blue block. These pairwise interaction features, $I_{1,1},\ldots I_{p,q}$, are averaged over all pairs (see Figure \ref{fig:architecture-all}: Mean Pooling) and passed through the Final Dense Layer (see Figure \ref{fig:architecture-all}), which consists of a single dense neural network layer, to obtain a single scalar prediction. The final prediction of the network is the minimum of this single scalar prediction and the Glide VDW score. 

\newpage

\newpage
\section{Supplementary Figures} \label{sup:figures}

\begin{figure}[h]
\centering
\includegraphics[width=\textwidth]{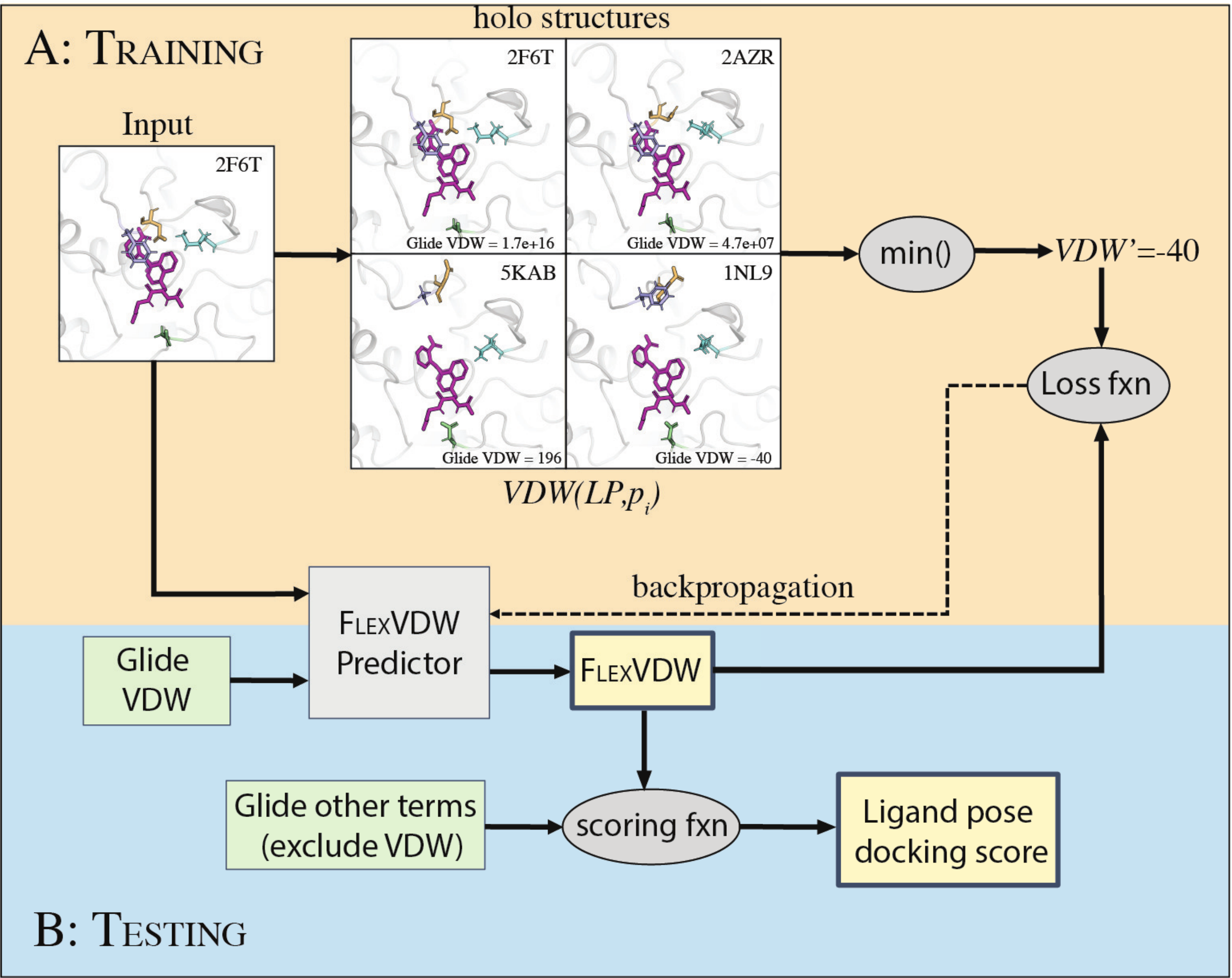}
\caption{High-level overview of training and testing of 
\label{supfig:pipeline}
\textsc{FlexVDW}. (A): Given a protein-ligand pose structure and its Glide VDW score, we train \textsc{FlexVDW} to predict the VDW score of the ligand pose accounting for the protein flexibility by using $VDW'$ as training label. $VDW'$ is defined as the minimum of the VDW scores of the ligand pose when superimposed on other holo structures of the same protein (see equation \ref{eq:min_vdw}). Here, the input ligand pose clashes with the loop residues PHE182 (light purple) and ASP181 (yellow) of the input docking structure 2F6T. When superimposed on other holo structures, the ligand pose severely clashes with the loop residues in structure 2AZR. This loop, which contains amino acid residues PHE182 and ASP181, is located further upwards in structures 5KAB and 1NL9 compared to in 2F6T and 2AZR. In structure 5KAB, there are only minor clashes of the ligand pose with with residues ASP48 (green) and LYS120 (cyan), and in structure 1NL9, there is no clash at all. The protein conformation in 1NL9 fits the ligand pose, giving a favorable (i.e., low) VDW score of $-40$, even though the VDW score of the ligand pose with respect to the input docking structure 2F6T is very large (it is 1.7e+16) due to severe clashes with the loop residues in the input docking structure. 
(B): During testing, \textsc{FlexVDW} takes only a single protein-ligand pose structure and Glide VDW score as input and outputs a predicted VDW score of the ligand pose with respect to the input protein structure that takes into account implicit protein flexibility. The predicted VDW score is integrated with other Glide terms (excluding the Glide VDW score) to calculate the final docking score of the ligand pose.} 
\end{figure}

\begin{figure}[h]
\centering
\includegraphics[width=1\textwidth]{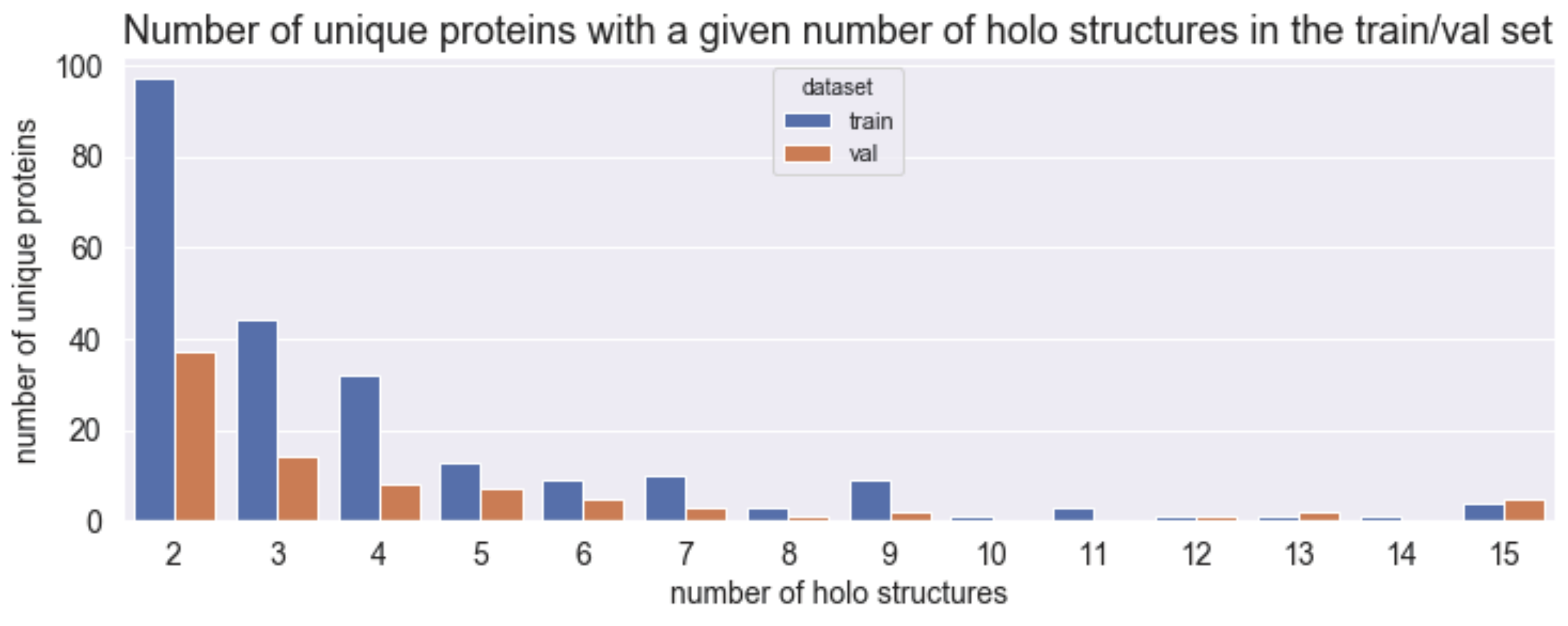}
\caption{Distribution of the number of holo structures per unique protein used to generate the labels and docking poses in the training and validation sets. We limit the maximum number of holo structures to 15 to prevent overrepresentation of certain proteins.}
\label{supfig:holo_dist}
\end{figure}

\begin{figure}[h]
\centering
\includegraphics[width=1\textwidth]{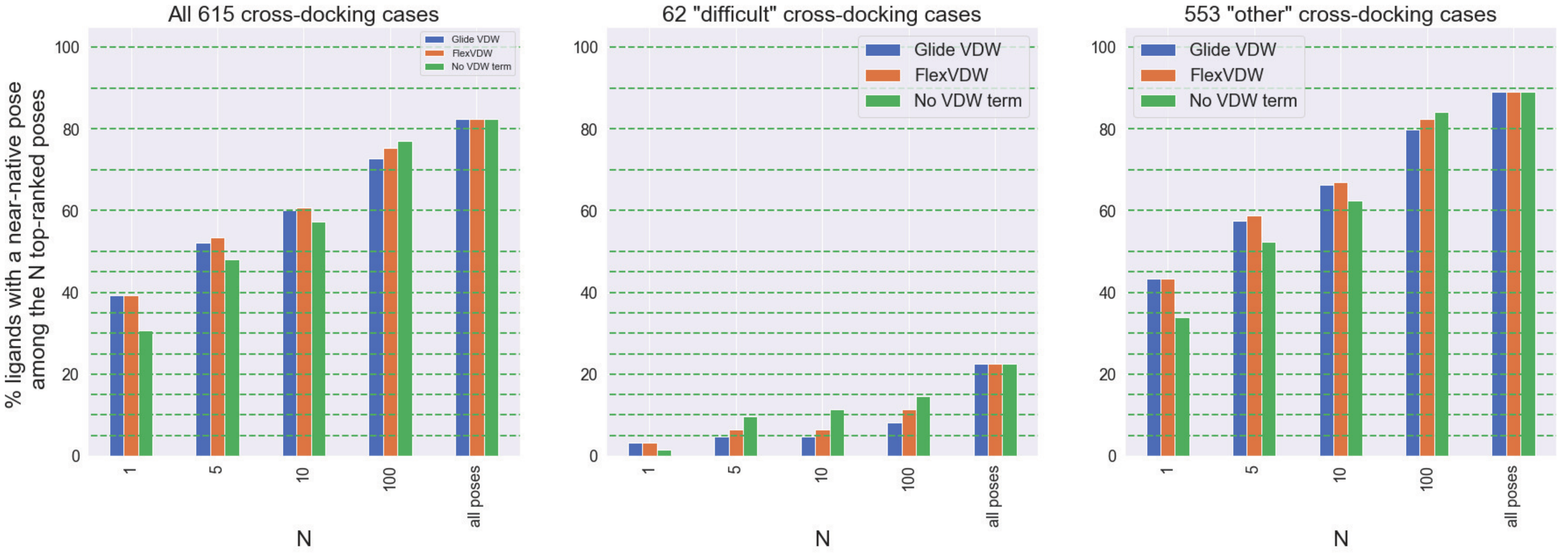}
\caption{Percentage of cases for which a near-native pose is included in the top-N poses sorted by docking score (higher is better). This analysis differs from that of Figure \ref{fig:ml_soft2} only in that the refined native ligand pose was not added to the set of poses considered for each ligand. The bars labeled "all poses" show the fraction of cross-docking pairs in which any near-native pose is present in the set of candidate poses. Only about 80\% of the cross-docking pairs have at least one near-native pose in the generated candidate poses, which makes the comparison between methods difficult, especially for the "difficult" cases, because for most of those cases, the candidate pose set does not include any near-native pose.}
\label{supfig:glide_top_n}
\end{figure}

\begin{figure}[h]
\centering
\includegraphics[width=1\textwidth]{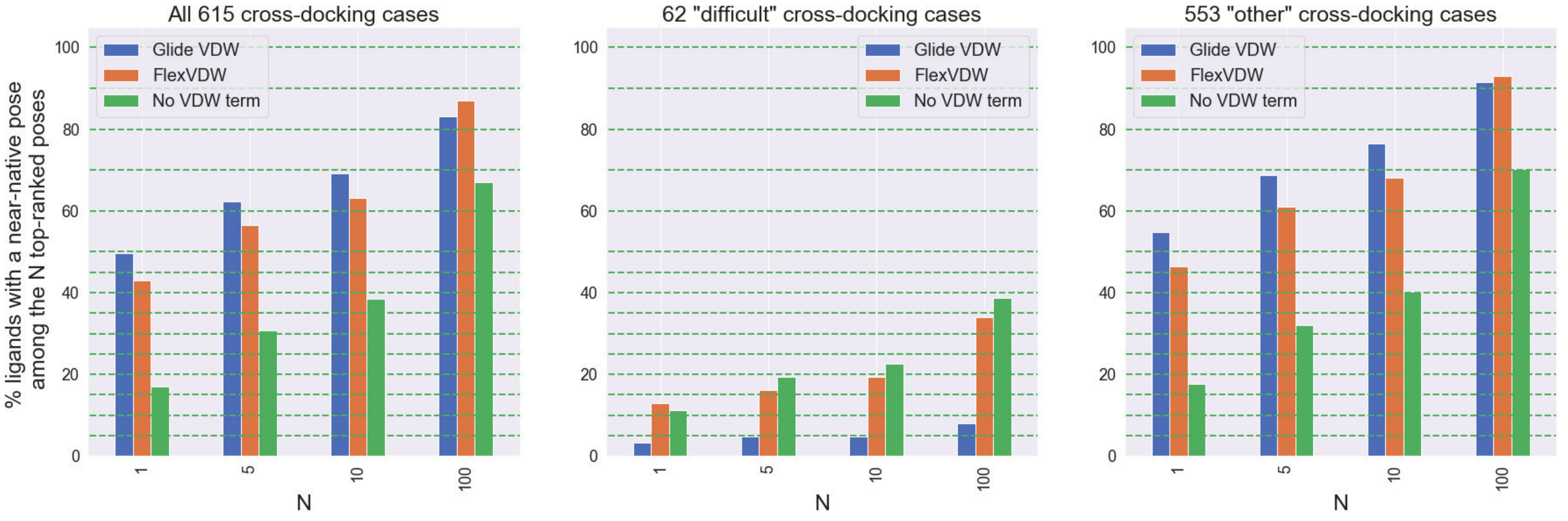}
\caption{Percentage of cases for which a near-native pose is included in the top-N poses sorted by docking score (higher is better). This analysis differs from that of Figure \ref{fig:ml_soft2} only in that we use VDW radius scaling of 0.0/0.0 instead of 0.6/0.5 for receptor/ligand for the softened VDW parameters. This further softening of VDW parameters leads to generation more "garbage" poses (i.e., poses with severe collisions with receptor backbones). As can be seen in the figure, the performance of \textsc{FlexVDW} is worse here compared to in Figure \ref{fig:ml_soft2}. This is to be expected since the ML model never sees such poses during training. However, a scoring function that includes \textsc{FlexVDW} generalizes better than one that does not include a VDW term.}
\label{supfig:ml_softest}
\end{figure}

\begin{figure}[h]
\centering
\includegraphics[width=0.65\textwidth]{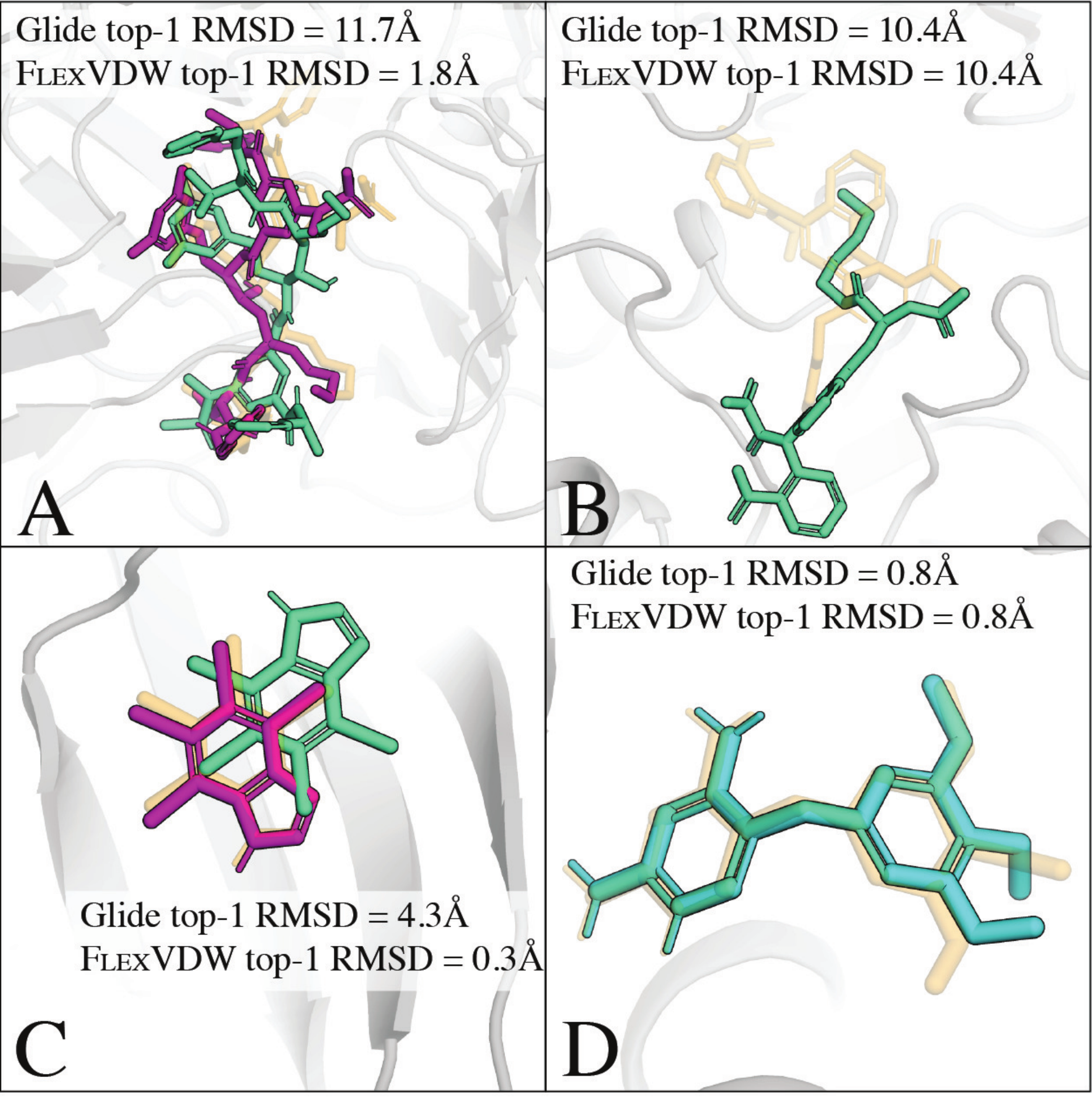}
\caption{Comparison of the posed ranked first (top-1) by Glide (cyan) and \textsc{FlexVDW} (magenta). The experimentally determined native poses are shown in transparent light orange. (A): Ligand of 3I25 docked to the protein structure of 1FKN (Uniprot ID: P56817). Due to clashes of several residue side chains with the native ligand pose (see Figure \ref{fig:ml_vs_glide_vdw}A), Glide’s top-1 ligand pose (cyan) was placed in the opposite orientation to the native ligand pose. \textsc{FlexVDW}, on the other hand, was able to select a near-native ligand pose as the top-1 pose. (B): Ligand of 1N06 docked to the protein structure of 1BZC (Uniprot ID: P18031). Although \textsc{FlexVDW} predicts a negative VDW score for the native ligand pose (see Figure \ref{fig:ml_vs_glide_vdw}B), the near-native poses are eventually rejected due to the high electrostatic repulsion energy (not shown in the figure). Hence, \textsc{FlexVDW} selects the same pose as Glide as top-1. (C): Ligand of 2OXY docked to the protein structure of 2PVJ (Uniprot ID: P28523). \textsc{FlexVDW} selects a near-native pose as top-1, while Glide does not. (D): Ligand of 2W3A is docked to the protein structure of 1BOZ (Uniprot ID: P00374). Both Glide and \textsc{FlexVDW} select the same pose as top-1.}
\label{supfig:ml_vs_glide_top_1}
\end{figure}


\end{document}